\documentclass[sigconf]{acmart}
\renewcommand\footnotetextcopyrightpermission[1]{}
\usepackage{booktabs} 
\usepackage{balance}

\setcopyright{rightsretained}

\acmDOI{}

\acmISBN{}

\acmConference[VAMS'17]{ACM VAMS Workshop}{August 2017}{Como, Italy} 
\acmYear{2017}
\copyrightyear{2017}

\acmPrice{15.00}

\begin{document}
\title{Patterns of Multistakeholder Recommendation}

\author{Robin Burke}
\orcid{0000-0001-5766-6434}
\affiliation{%
  \institution{School of Computing}
  \streetaddress{DePaul University}
  \city{Chicago} 
  \state{Illinois} 
}
\email{rburke@cs.depaul.edu}

\author{Himan Abdollahpouri}
\orcid{}
\affiliation{%
  \institution{School of Computing}
  \streetaddress{DePaul University}
  \city{Chicago} 
  \state{Illinois} 
}
\email{habdolla@depaul.edu}

\begin{abstract}
Recommender systems are personalized information systems. However, in many settings, the end-user of the recommendations is not the only party whose needs must be represented in recommendation generation. Incorporating this insight gives rise to the notion of multistakeholder recommendation, in which the interests of multiple parties are represented in recommendation algorithms and evaluation. In this paper, we identify patterns of stakeholder utility that characterize different multistakeholder recommendation applications, and provide a taxonomy of the different possible systems, only some of which have currently been implemented.
\end{abstract}


\maketitle

\section{Introduction}

\noindent Recommender systems provide personalized information access, supporting e-commerce, social media and other applications where the volume of content would otherwise be overwhelming for users. They have become indispensable tools of the Internet age, found in systems of many kinds. 

One of the defining characteristics of recommender systems is personalization. Recommender systems are typically evaluated on their ability to provide items that satisfy the needs and interests of the end user, and to do so in a way that recognizes the unique tastes and needs of each. However, there is increasing recognition that the receiver of the recommendation is not the sole party of interest in recommendation generation. 

This phenomenon was first recognized in person-to-person recommendation, where the recommendations are pointers to other users, as in online dating. In such \textit{reciprocal recommendation} scenarios, it makes sense to recommend person B to person A, only if it is likely that person B will also be interested in making that connection~\cite{reciprocal,reciprocaldating}. Otherwise, it would be a waste of person A's time and attention to receive such a recommendation. Reciprocal recommendation arises naturally in a number of other contexts, including job seeking~\cite{rodriguez_multiple_2012}, peer-to-peer ``sharing economy'' recommendation (such as AirBnB, Uber and others), on-line advertising \cite{targetadvertisingbiding}, and scientific collaboration~\cite{lopes2010collaboration,tang2012cross}.

To these considerations, we also add the concerns of the platform itself. In e-commerce settings, the recommender system is often seen as an extension of a site's marketing function, and, as such, business considerations may enter into the generation and display of recommendations. In other settings, such as employment or housing, equity considerations may enter into the provision of recommendations to users, making it desirable to achieve fair recommendation outcomes~\cite{DBLP:journals/corr/YaoH17,burke_robin_multisided_nodate}. For example, an educational recommender may have pedagogical goals in linking students with experiences, and these goals may be quite different from what students would choose for themselves~\cite{burke_educational_2016}.

Thus, we have argued that recommender systems research needs to adopt a multistakeholder perspective on system design and evaluation: recognizing the needs and concerns of multiple parties involved in recommendation outcomes~\cite{abdollahpouri_recommender_2017}. The aim of this paper is to investigate the different types of multistakeholder recommender systems that may arise, given different requirements coming from different stakeholders, and to provide a taxonomy to describe existing systems and to guide future research.

\section{Prior research}

\noindent The concept of multistakeholder recommendation builds on three existing research themes in the recommender systems community and beyond: (1) the economics of multisided platforms; (2) the growing interest in multiple objectives for recommender systems, including such concerns as diversity and serendipity; and, (3) the application of personalization to specific matching problems, as in reciprocal recommendation described above.

\subsection{Multisided platforms}

\noindent One way to understand the multistakeholder aspect of recommendation is through the recent emergence of the theory of multisided platforms~\cite{rochet2003platform,evans_platform_2011}. A multisided platform is a system that brings together individuals participating in a given transaction, reducing their transactions costs as to make the interaction more profitable for those involved. A business that creates such a platform can profit from it, even though, in theory, the parties could engage in such transactions on their own. 

An example, from~\cite{evans_matchmakers:_2016}, is the online restaurant reservation site OpenTable\footnote{www.opentable.com}. It is a two-sided platform for users to find restaurants with available reservations and restaurants to present themselves to users looking for dining spots. OpenTable doesn't offer collaborative recommendations at present\footnote{By way of recommendations, the site offers a listing of restaurants where the user has previously booked a reservation.}, but it is easy to envision how such recommendations could be seen as a simple extension to the matchmaking function that the site performs. 

Adopting a multisided view of the platform in which a recommender system is embedded leads naturally to the idea that recommender systems themselves must take multiple perspectives into account in order to be successful. This is not a traditional position on recommender system evaluation -- which has typically focused on user-oriented objectives, such as accuracy~\cite{Herlocker04evaluatingcollaborative}. It could be argued, however, that industrial recommender systems research, with its use of customer lifetime value and other ``key performance indicators'', has already begun to shift to attention away from a strict focus on the customer's needs and preferences~\cite{agarwal2015statistical}. 

One key finding from multisided platform research is that these platforms are extremely diverse and their structure is governed by the characteristics of the specific business niche that a platform fills~\cite{evans_matchmakers:_2016}. Such domain-specific considerations are well-understood in recommender systems research. 

Because of our emphasis on the stakeholders on multiple sides of a transaction, we are excluding issues of group recommendation from our discussion in this paper. A group recommender provides recommendations to a set of consumers, all of whom are on the same side of the transaction. There are important research issues in providing recommendations in such contexts, but the multistakeholder considerations we discuss here are less applicable since the entire group is on the receiving side of the same recommendation interaction. 

\subsection{Multiobjective recommendation}

\noindent As interest has grown both in reciprocal recommendation and the integration of business considerations in recommendation, there has been an increased level of research into the use of multiple objectives in recommendation, including \cite{jambor_optimizing_2010,agarwal_click_2011,svore_learning_2011,rodriguez_multiple_2012,jiang_optimization_2012,agarwal_personalized_2012,biasRecSys2017}. These techniques provide a way to limit the expected loss on one metric (typically accuracy) while optimizing for another, such as diversity or coverage. 

While most multiobjective recommendation research concentrates on maximizing objectives for the end user alone, there are some examples of systems that incorporate objectives from multiple stakeholders. For example, \cite{rodriguez_multiple_2012} combines the perspectives of recruiters and users at LinkedIn and shows that an integrated objective yields more fruitful interactions. Research in social network settings has noted that there may be general, system-wide, benefits to other stakeholders in both person and media content recommendation, since actions taken by one user may impact the network as a whole~\cite{daly2010network,ronen2014recommending}.

Another area of recommendation that explicitly takes a multiobjective perspective is the area of health and lifestyle recommendation. Multiple objectives arise in this area because users' short-term preferences and their long-term well-being may be in conflict~\cite{lin2011motivate, ponce2015quefaire}. In such systems, it is important not to recommend items that are too distant from the user's preferences -- even if they would maximize health: for example, a food recommender's ideal low-fat diet might be too extreme for a user accustomed to heartier fare. The goal to be persuasive requires that the user's immediate context and preferences be honored~\cite{ge2015health}. 

\subsection{Personalization for matching} 

\noindent The problem of matching multiple parties under preference constraints has a long history in the economics literature. See, for example, the work of Roth ~\cite{roth1982economics,roth1992two,roth2015gets}. These studies assume that the  algorithm makes all of the matches at once, a setting incompatible with online systems. The BALANCE algorithm relaxes this requirement, implementing online matching in the presence of budgetary constraints~\cite{optimalmatchingKalyanasundaram}. However, a system that outputs a stable matching is quite different from one that outputs a personalized ranking and where we cannot assume that the system has full knowledge of the user's preferences.

The concept of multiple stakeholders in recommender systems is suggested in a number of prior research works. As discussed above, researchers in reciprocal recommendation appliations such as job search and online dating, have looked at bi-lateral considerations to ensure that a recommendation is acceptable to both parties ~\cite{reciprocaldating}. Similar ideas have appeared in group recommender systems where the goal is to find recommendation(s) that can maximize the utility of all users in the group \cite{masthoff2011group}. 

The field of computational advertising has given considerable attention to balancing personalization with multistakeholder concerns. Auctions, a well-established technique for balancing the interests of multiple agents in a competitive environment, are widely used both for search and display advertising \cite{internetadvertisingyuan}. For example, the AdWords system within the Google search engine ranks bidders by the product of their bid and a system-determined ``quality score'', a ranking intended to maximize the expected revenue for the search engine~\cite{mehta2007adwords}. 

Advertisers also bid for the opportunity to show display advertising to web users in display auctions~\cite{yuan2013real}. There is a substantial literature in real-time targeted advertising in which advertisers' expected revenue and / or available budget are incorporated into the decision to deliver personalized advertising to a user. See \cite{optimalbiding} for an example.

In these advertising settings, the amount of information about each user is generally quite limited, and privacy considerations control how much of it can be shared with advertisers. In addition, the real-time nature of the application and the huge potential user base makes user-level personalization computationally impractical. As a result, ad display is personalized only to a very coarse degree (age and geographic region, for example), if at all. 

\section{A Multistakeholder Taxonomy}
We have argued that a comprehensive approach to multistakeholder recommendation should explicitly represent the different stakeholders in the recommendation process and formalize their utilities~\cite{soappaper,burke_educational_2016}. A utility-oriented approach is sufficiently general that a wide variety of recommendation scenarios can be represented including reciprocal recommendation, online budget management, and others.

One important finding in the economics of multisided platforms is that different applications require different distributions of utility. In many multisided platforms, there is a ``subsidy side'' of the transaction where one set of parties uses the platform at a reduced cost or no cost~\cite{evans_matchmakers:_2016}. For example, users of OpenTable do not pay to make reservations; instead, restaurants pay for each reservation made.

In recommender systems as well, the outcomes of the system may be biased towards one group of parties for similar reasons. In addition, the need for personalization may vary across systems. For example, in an e-commerce site, product suppliers will likely not care about the characteristics of consumers: as long as products are recommended to likely buyers, they will be satisfied with the behavior of the system. Online advertising is different: typically an ad campaign is targeted towards a particular audience. Therefore, a recommended ad placement is only successful if the ad matches the user's interests, to the extent they are known, and the user matches the target audience towards which the campaign is oriented. 

In addition to the consumers and providers, the system itself may or may not have preferences for certain types of matches produced by recommendations. As a diverse marketplace with a variety of sellers, the artisan-focused site Etsy is interested in attracting and keeping  sellers as well as buyers. Etsy spends a great deal of effort supporting and marketing to its sellers\footnote{Etsy's seller site can be found at  https://www.etsy.com/sell}. In such an environment, nurturing new entrants may be a platform owner priority, and this consideration might enter into the utility the owner associates with a set of recommendations, leading it, for example, to prefer recommendation lists that contain a mix of new and established sellers over those with only established ones.

We divide the stakeholders of a given recommender system into three categories: consumers $C$, providers $P$, and platform or system $S$. The consumers are those who receive the recommendations. They are the individuals whose choice or search problems bring them to the platform, and who expect recommendations to satisfy those needs. The providers are those entities that supply or otherwise stand behind the recommended objects, and gain utility from the consumer's choice. The final category is the platform itself, which has created the recommender system in order to match consumers with providers and has some means of gaining benefit from doing so.  

\subsection{Incorporating utilities}
We can classify recommender systems based on how they support and incorporate the utilities of these stakeholders. A system may support \textit{passive} or \textit{active} interaction (or both). A passive system provides recommendations unprompted by the user, such as ``Recommended for you'' items that are listed on many sites. Active systems provide recommended items in response to some type of query or interaction from the user. 

We can extend this notion of active vs. passive to providers as well. A system may allow providers to take an active part in specifying target consumers, or it may not. In our notation, we will represent a passive approach with the superscript $-$ and an active one with the superscript $+$.

Another distinction has to do with the type of personalization an application supports. We can characterize personalization in terms of stakeholder utility. If the system does not attempt to provide items that maximize a given stakeholder's utility, we say that the system is neutral with respect to that stakeholder, and use the subscript $n$. For example, an e-commerce site may not consider the provider when delivering recommendations. If the system is neutral with respect to the user, we might not be willing to call it a recommender system, but we can consider it a degenerate case of non-personalized recommendations -- the list of ``Most popular'' or ``Trending'' items on some sites is an example.

The system may attempt to satisfy the needs of the user by offering recommendations tailored (at least in part) to the preferences of that stakeholder. This condition is denoted with the subscript $p$. When the stakeholder is the consumer, this is the most common personalized recommendation scenario. When there is personalization for both the consumer and the provider, we have the reciprocal recommendation scenario described above.

\begin{figure*}[tbh]
	\centering
	\includegraphics[width=10cm]{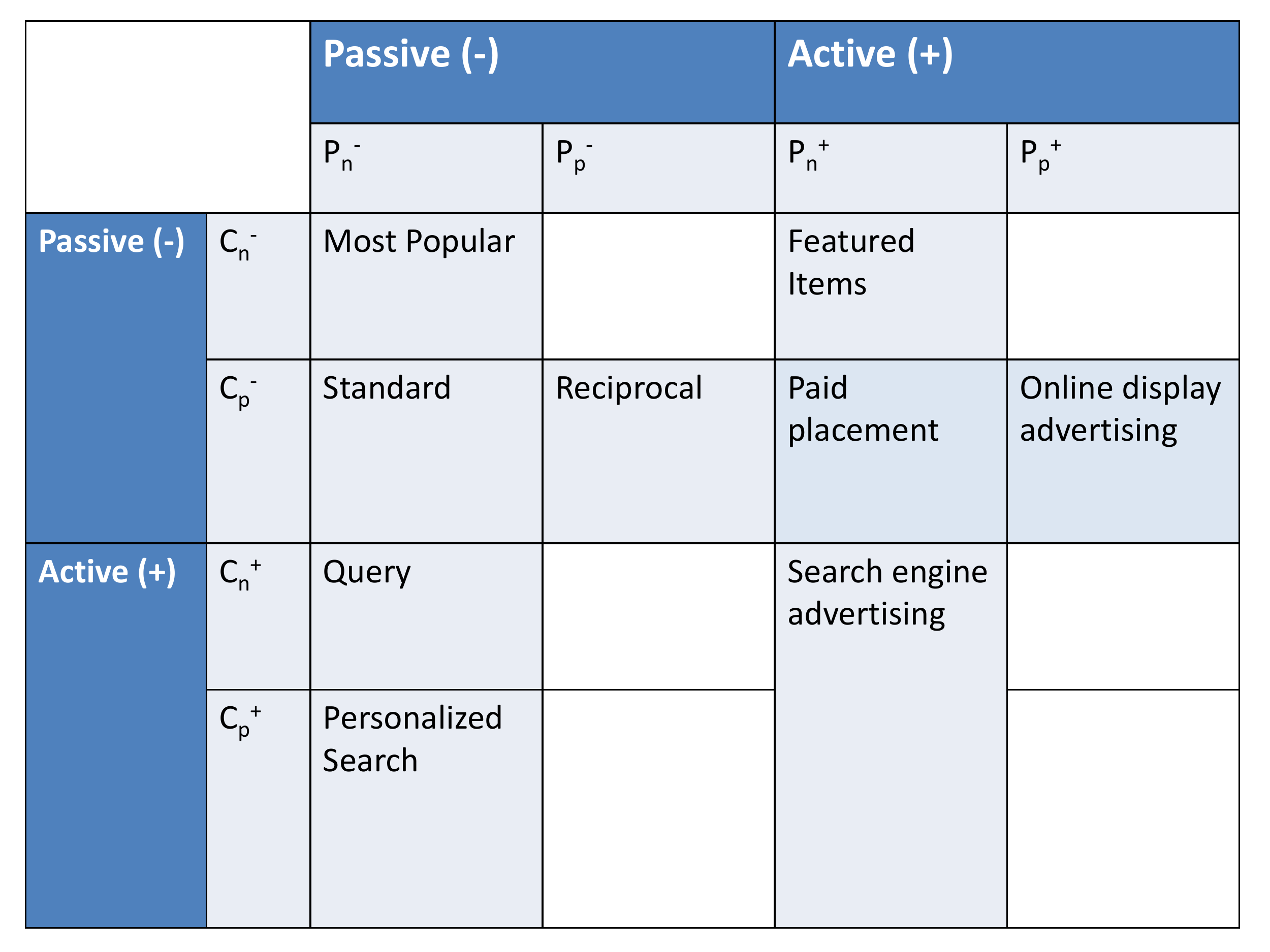}
	\caption{Platform-neutral multistakeholder designs. White areas are unexplored. \label{fig:extant}}
\end{figure*}

\subsection{Consumer and provider utilities}\label{sec:utilities}
Figure~\ref{fig:extant} shows a taxonomy of different combinations of consumer and provider utility configurations with examples of unexplored configurations blank. The rows enumerate the different possibilities for consumers: passive vs active, and neutral vs personalized. The columns enumerate the same possibilities for providers. We can identify each entry with a unique tuple: $\langle C, P \rangle$ with the appropriate subscript and superscript.

A passive system that provides items without regard for the utilities of the user or provider would be a simple ``Most Popular'' list: $\langle C_n^-, P_n^- \rangle$. Adding consumer personalization, we get the standard e-commerce recommendation scenario, $\langle C_p^-, P_n^- \rangle$. If the user formulates a query or other prompt to the system, we have an active system that can also be neutral: just returning answers to the query, or one that personalizes the search results, as is more and more common in search engines. These options are represented by tuples $\langle C_n^+, P_n^- \rangle$ and $\langle C_p^+, P_n^- \rangle$

If the system is not neutral with respect to the provider, it must try to match consumers to provider preferences. Such designs are less common in research and in practice, as indicated by the blank areas in Figure~\ref{fig:extant}. The typical reciprocal recommendation system is characterized by the tuple $\langle C_p^-, P_p^- \rangle$. Here, provider preferences are gathered through provider actions, for example, accepting offers made by recommendation consumers -- such as an AirBnB host deciding whether or not to accept a guest.

In the active case, the provider can specify the type of consumer that is a desirable target. If we do not care about the consumers' preference, we can offer a list of ``Featured items', selected based on the providers' audience choices: $\langle C_n^-, P_p^- \rangle$. ``Promoted posts'' within social media sites are a good example of the second case ($\langle C_n^-, P_p^+ \rangle$) -- these are recommendations given to users for whom they are a good fit, but only if the provider has paid to appeal to those targeted consumers. 

Where users are actively posing a query, we see familiar scenarios from search engine advertising. The placement of such ads also takes place through a bidding process in which the providers' bids and the click likelihood are scored~\cite{edelman2007internet}. To the extent that click likelihood is calculated on a per-user basis, such systems could fall into either of the active categories, $\langle C_n^+, P_n^+ \rangle$ or $\langle C_p^+, P_n^+ \rangle$\footnote{In some settings, the advertiser may use geo-targeting or otherwise identify the characteristics of users in their target market. Such cases would be (somewhat weak) examples of the $P_p^+$ category.}.

The only case that is arguably personalized with respect to the provider's utility is found in online display advertising. Here there is a form of reciprocation in that the ad should be appealing to the user and the user should be in the defined target audience. Typically, a vector of demographic features are available on which basis the targeting decision can be made. Thus, the tuple $\langle C_p^-, P_p^+ \rangle$ would apply. Real-time auctions are used in which providers are expected to quantify the utility they expect from a given ad placement~\cite{internetadvertisingyuan}. 

\subsection{The platform as a stakeholder}
As discussed above and consistent with the economics of multiplatform systems, we believe it is important to add a third dimension that takes into account the utility of the platform itself. To do so, we need a 3-tuple to represent the possible system types: $\langle C, P, S \rangle$. 

All of the examples discussed in Section~\ref{sec:utilities} are ones in which the system is neutral: $S_n$, neither gaining or losing utility based on what results are produced or to whom they are shown. This is the classical model of how recommender systems operate. However, in many real-world contexts, the system may have some aggregate utility that is factored into recommendation decisions. Consistent with our notation above, we will denote this possibility with the notation $S_a$, for aggregate. For example, the system may prefer to recommend high-margin items, all other considerations being equal, with the aim of having higher average profit per user visit. Recent research in systems with such $\langle C_p^-, P_n^-, S_a \rangle$ designs include \cite{moveRecProfitMax,persolizedPromotion,jiang_optimization_2012}. A search-oriented example is Google's ``quality score'' discussed above in which the search engine takes the expected revenue into account in deciding what ad to show, a system that fits into the $\langle C_{n,p}^+, P_n^+, S_a \rangle$ category. 

Alternatively, the system may seek to tailor outcomes specifically to achieve particular objectives separate from stakeholder utilities -- a situation that we denote with $S_t$, for targeted. For example, an educational site may view the recommendation of learning activities as a curricular decision and seek to have its recommendations fit a model of student growth and development~\cite{burke_educational_2016}. Its utility may therefore go beyond a simple aggregation of those of the other stakeholders. This problem also arises when fairness is a consideration in recommendation~\cite{burke_robin_multisided_nodate}. For example, consider a recommender system suggesting job opportunities to job seekers. An operator of such a system might wish, for example, to ensure that male and female users with similar qualifications get recommendations of jobs with similar rank and salary. Such an objective goes beyond a simple aggregation of job seekers' utilities. 

\section{Conclusion}

These considerations leave us with a complex picture of the multistakeholder recommendation space. There are two basic forms of personalization: neutral $n$ and personalized $p$ results. These can be exercised in both passive $-$ and active $+$ recommendation contexts, and can be applied to both consumers and providers. As shown in Figure~\ref{fig:extant}, this gives a total of 16 possibilities. If we are interested in personalized recommendations, the $C_n$ cases are of less interest, leaving 8 designs, of which really only two have seen substantial attention in recommender systems research: $\langle C_p^-, P_n^-, S_n \rangle$ and $\langle C_p^-, P_p^-, S_n \rangle$. 

Furthermore, as we have shown, the platform itself is a stakeholder. Many platforms are neutral with respect to the recommendation function, with the result that the $S_n$ condition is most fully explored. However, there is an increasing trend to represent the system's own utility in some fashion, typically oriented towards profit or other performance indicators. Thus, we expect to see additional exploration of the $S_a$ space in both research and application. Finally, there are emerging recommendation research agendas in health, education, and other areas where the platform has its own strategic considerations to guide recommendation activity and this is where $S_t$ designs are beginning to be seen. 

Thus, the full set of possibilities for multistakeholder recommender system design is a three-dimensional version of Figure~\ref{fig:extant}, a 2x4x3 matrix of possibilities (excluding the non-personalized options), which, except for the $S_n$ subset, has been only lightly explored. We look forward to future research that will fill these gaps.

\section{Acknowledgments}
\noindent This work is supported in part by the National Science Foundation under grant IIS-1423368.

\bibliographystyle{ACM-Reference-Format}
\balance
\bibliography{myref.bib} 

\end{document}